\documentclass[preprint,aps,preprintnumbers,showpacs,showkeys,superscriptaddress]{revtex4-2}
\usepackage{graphicx}  
\usepackage{dcolumn}   
\usepackage{bm}        
\usepackage{amssymb}   
\usepackage{subfigure}
\usepackage{epstopdf}
\usepackage{color}
\usepackage[utf8]{inputenc}
\usepackage{amsmath}
\usepackage{array}
\usepackage[dvipsnames]{xcolor}
\usepackage{graphicx}
\usepackage{amsmath}
\usepackage[T1]{fontenc}
\usepackage{newtxtext, newtxmath}
\usepackage{gensymb}
\usepackage{makecell}
\usepackage{caption}
\usepackage{listings}
\usepackage{multirow}
\usepackage{hhline}
\usepackage{subcaption}
\usepackage{enumitem}
\usepackage{tabularx}
\usepackage{hyperref}
\usepackage{float}

\allowdisplaybreaks[4]

\begin{document}

\title{Sensitivity Study of the Tau Lepton Electric Dipole Moment at the Super Tau-Charm Facility}
\thanks{This work is supported by the National Key R\&D Program of China (2022YFA1602200, 2023YFA1607200); the National Natural Science Foundation of China (NSFC) (12341501, 12341503, 12341504); the international partnership program of the Chinese Academy of Sciences (211134KYSB20200057). Xulei Sun is supported by the Undergraduate Research Program of the University of Science and Technology of China and the Xinhe Scholar Program of the School of the Gifted Young.}
\author{Xulei Sun\href{https://orcid.org/0009-0002-1709-9302}{\includegraphics[scale=0.1]{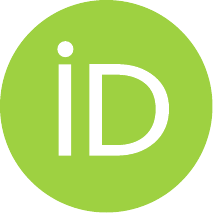}}}
\email{sunxulei@mail.ustc.edu.cn}
\affiliation{School of the Gifted Young, University of Science and Technology of China, Hefei 230026, China}
\author{Yongcheng Wu\href{https://orcid.org/0000-0002-1835-7660}{\includegraphics[scale=0.1]{figure/logo-orcid.pdf}}}
\email{ycwu@njnu.edu.cn}
\affiliation{Department of Physics, Institute of Theoretical Physics and Institute of Physics Frontiers and Interdisciplinary Sciences, Nanjing Normal University, Nanjing, 210023, China}
\affiliation{Nanjing Key Laboratory of Particle Physics and Astrophysics, Nanjing, 210023, China}
\author{Xiaorong Zhou\href{https://orcid.org/0000-0002-7671-7644}{\includegraphics[scale=0.1]{figure/logo-orcid.pdf}}}
\email{zxrong@ustc.edu.cn}
\affiliation{Department of Modern Physics, School of Physics, University of Science and Technology of China, Hefei 230026, China}

\begin{abstract}
This study investigates the intrinsic electric dipole moment (EDM) of the $\tau$ lepton, which is an important quantity in the search for physics beyond the Standard Model (BSM). In preparation for future measurements at the Super Tau-Charm Facility (STCF), we employ Monte Carlo simulations of the $e^+e^- \rightarrow \tau^+\tau^-$ process and optimize the analysis methodology for EDM extraction. Machine learning techniques are implemented to efficiently identify signal events ($\tau^\pm\rightarrow\pi^\pm\pi^0\nu_\tau$), which result in a significant improvement in signal-to-noise ratio. Our optimized event selection algorithm achieves $80.0\%$ signal purity with $6.3\%$ efficiency. We develop an analytical approach for $\tau$ lepton momentum reconstruction and derive the squared spin density matrix along with optimal observables, which maximize the sensitivity to $d_\tau$. The relationship between these observables and the EDM is established with the estimated sensitivity of $|d_\tau| < 3.89\times 10^{-18}\,e\cdot\mathrm{cm}$ at a $68\%$ confidence level. These results provide a foundation for future experimental measurements of the $\tau$ lepton EDM in STCF experiments.
\end{abstract}

\maketitle
\flushbottom
\clearpage

\section{Introduction}
The electric dipole moment (EDM) is a physical quantity that characterizes the distribution of electric charge within a system. Theoretically, the EDM of a point-like particle is expected to be zero. Within the Standard Model (SM), due to charge-parity-conjugation (CP) violation, the intrinsic EDM of the $\tau$ lepton is predicted to be approximately ${10}^{-37}e\,\mathrm{cm}$~\cite{CPViolation,SM,BSM}, far below the experimental sensitivity. From the perspective of symmetry, a non-zero EDM signifies CP violation, which is a crucial prerequisite for the observed matter-antimatter asymmetry in the current observable universe. The CP violation predicted by the SM is minute and insufficient to account for the observed asymmetry~\cite{CPViolation2}. However, some Beyond the Standard Model (BSM) theories predict a significantly larger value, around ${10}^{-19}e\,\mathrm{cm}$~\cite{CPViolationTau1,CPViolationTau2}, making it feasible to measure the $\tau$ EDM experimentally and potentially reveal new physics.

For stable particles such as neutrons and electrons, EDM measurements typically employ precise experimental methods based on the spin precession phenomenon in strong electric or magnetic fields~\cite{Precession1,Precession2}. When a particle with a non-zero EDM interacts with an electric field, its spin direction will undergo a slight deflection, which can be detected by highly sensitive experimental apparatuses. To date, numerous experiments have set extremely stringent upper limits on the EDMs of neutrons and electrons, further corroborating the predictions of the SM while also providing crucial experimental constraints for new physics searches beyond the SM.

For short-lived particles such as heavy quarks and leptons, direct EDM measurements face significant challenges due to their extremely short lifetimes, which generally preclude traditional spin precession experiments~\cite{UnstableParticle}. Consequently, indirect approaches are adopted, wherein deviations in scattering cross-sections or decay rates from SM predictions are used to search for potential EDM signals of these particles. This approach not only broadens the applicability of EDM measurements but also opens new experimental avenues for exploring BSM physics~\cite{EDMforBSD}.

The most precise measurement of the $\tau$ EDM to date comes from the Belle experiment in Japan, which studied the process $e^+e^-\rightarrow\tau^+\tau^-$ at the KEKB collider, setting an upper limit of ${10}^{-17}e\,\mathrm{cm}$ on the $\tau$ EDM~\cite{Belle2022,UsingBelle}. To further explore the potential of measuring the $\tau$ EDM in the tau-charm energy region, we consider the next-generation electron-position collider, the Super Tau-Charm Facility (STCF), which is under design and construction in China. The STCF is designed to operate in the energy range of $2.0-7.0\,\mathrm{GeV}$ with a peak luminosity of $0.5\times{10}^{35}\,\mathrm{cm^{-2}s^{-1}}$, allowing an integrated luminosity of $1\,\mathrm{ab^{-1}}$ per year. At center-of-mass energies (CMEs) of $4.2\,\mathrm{GeV}$ and $7.0\,\mathrm{GeV}$, the STCF is expected to produce $3.5\times{10}^9$ and $1.7\times{10}^9$ $\tau$ pairs per year, respectively, significantly enhancing the sensitivity of $\tau$ EDM measurements. Additionally, the STCF will utilize a vertex detector to improve vertex resolution~\cite{STCF}. Compared to the high-energy LEP experiments and the $10\,\mathrm{GeV}$ Belle experiment, the tau-charm energy region offers reduced radiative $\tau$ pair events, higher $\tau$ pair production cross-sections, and more accurate charged particle identification and photon reconstruction, leading to a higher reconstruction efficiency for $\tau$ leptons.

The Lagrangian related to the EDM of $\tau$ lepton is given by~\cite{Belle2003,Wu}:
\begin{equation}
  \mathcal{L}_{\mathrm{CP}}=-i d_{\tau}\bar{\tau}\sigma_{\mu\nu}\gamma_{5}\tau \partial_\mu A_\nu.
\end{equation}
{Including the EDM term, the scattering amplitude for 
the tau-pair production at electron-positron collider $e^{+}(p_1) e^{-}(p_2) \rightarrow \tau^{+}\left(p_3,s_+\right) \tau^{-}\left(p_4,s_-\right)$ is given by:
\begin{align}
\mathcal{M}_{\rm prod} = \mathcal{M}_{\rm SM} + {\rm Re}(d_\tau)v\mathcal{M}_{\rm Re},
\end{align}
where $\mathcal{M}_{\rm SM}$ is the contribution from the SM without the EDM and $\mathcal{M}_{\rm Re}$ is the contribution from the EDM operator. {Note that we introduce extra $v$ in the second term in order to balance the dimensionality of $\mathcal{M}_{\rm SM}$ and $\mathcal{M}_{\rm Re}$. For simplicity}, we only keep the real part of ${\rm Re}(d_\tau)$. Then the squared matrix element contains three contributions:
\begin{align}
|\mathcal{M}_{\rm prod}|^2 &= |\mathcal{M}_{\rm SM}|^2+{\rm Re}(d_\tau)v|\mathcal{M}_{\rm inter}|^2+({\rm Re}(d_\tau)v)^2|\mathcal{M}_{\rm Re}|^2,
\end{align}
where $|\mathcal{M}_{\rm inter}|^2=2{\rm Re}(\mathcal{M}_{\rm SM}^*\mathcal{M}_{\rm Re})$ is the interference between the SM contribution and the EDM operator and will be the key component for retrieving the information about the EDM of $\tau$ lepton.  Each term in the matrix element depends on the spin of $\tau^+$ and $\tau^-$:
\begin{align}
|\mathcal{M}_i|^2 = \overline{|\mathcal{M}_{i}|^2} (1 + h^+_{i\mu} s_+^\mu + h^-_{i\mu}s_-^\mu + c_{i\mu\nu}s_+^\mu s_-^\nu )\quad (i={\rm SM, inter, Re}).
\end{align}

The optimal observable~\cite{OptimalObservable} is used to maximize the sensitivity to $d_\tau$ which is given by:
\begin{equation}
\mathcal{O}_{\operatorname{Re}}=\frac{|\mathcal{M}_{\rm inter}|^2}{|\mathcal{M}_{\rm SM}|^2}, 
\label{OO}
\end{equation}
The mean value of the observable $\mathcal{O}_{\mathrm{Re}}$ is given by:
\begin{equation}
\langle\mathcal{O}_{\mathrm{Re}}\rangle\propto\int\mathcal{O}_{\mathrm{Re}}\mathcal{M}_{\mathrm{prod}}^{2}d\Pi\approx\int\mathcal{M}_{\mathrm{inter}}^2 d\Pi+\mathrm{Re}(d_\tau)v\int\frac{(|\mathcal{M}_{\rm inter}|^2)^2}{|\mathcal{M}_{\mathrm{SM}}|^2} d\Pi,
\end{equation}
where the integration is performed over the available phase space $\Pi$ and we ignore the higher order terms ($|\mathcal{M}_{\rm Re}|^2$). The mean value of the optimal observable is thus a linear function of ${\rm Re}(d_\tau)v$~\cite{Belle2022,Belle2003}:
\begin{align}
    \langle\mathcal{O}_{\rm Re}\rangle = a_{\rm Re}\cdot{\rm Re}(d_\tau)v + b_{\rm Re},
    \label{O_d}
\end{align}
where
\begin{equation}
a_{\mathrm{Re}}=\int\frac{(|\mathcal{M}_{\mathrm{inter}}|^2)^2}{|\mathcal{M}_{\mathrm{SM}}|^2}d\Pi, \quad b_{\mathrm{Re}}=\int|\mathcal{M}_{\mathrm{inter}}|^2 d\Pi.
\end{equation}
Therefore, the electric dipole moment $d_\tau$ can be obtained from $|\mathcal{M}_{\mathrm{SM}}|^2$ and $|\mathcal{M}_{\operatorname{inter}}|^2$ of which the computation depends on the $\tau$ lepton momenta and spin vectors, which will be further constructed from the momenta of the tau decay products.}

The $\rho\rho$ mode of $\tau$ pair production, $e^+e^-\rightarrow\tau^+\tau^-\left(\tau^+\rightarrow\pi^+\pi^0\bar{\nu}_\tau,\tau^-\rightarrow\pi^-\pi^0\nu_\tau,\pi^0\rightarrow \gamma\gamma\right)$, is the dominant channel and provides the most promising avenue for probing the intrinsic EDM of the $\tau$ lepton~\cite{Belle2022}. Consequently, this study focuses on this process as the signal process. By simulating the electron-positron annihilation process, we employ multivariate analysis to optimize selection criteria, filter signal events, and ensure a low background rate. We analyze the kinematic properties of the final state particles, compute the momentum and spin of the $\tau$ lepton, and obtain the optimal observables and their relationship with the EDM, facilitating the measurement of the $\tau$ EDM at the STCF.

\section{Event Selection}

{
\subsection{STCF detector system}
The STCF detector system is a sophisticated assembly designed to maximize physics potential in the $\tau$-charm energy region. From the interaction point outward, it features a tracking system comprising an inner tracker (ITK) using radiation-hard technologies like $\mu$RWELL-based gaseous detectors or MAPS-based silicon pixels, followed by a large main drift chamber (MDC) with helium-based gas for precise momentum measurement. Particle identification is achieved via a barrel RICH detector and an endcap time-of-flight system (DTOF), providing kaon-pion separation up to $2\,\mathrm{GeV/c}$. The electromagnetic calorimeter (EMC) uses pure CsI crystals for high-resolution photon detection and energy measurement. A superconducting solenoid generates a $1\,\mathrm{T}$ magnetic field for tracking, surrounded by an iron yoke for structural support and flux return. The outermost layer is a muon detector (MUD) combining resistive plate chambers and plastic scintillators for efficient muon identification. The system is optimized for high luminosity, with advanced data acquisition handling event rates up to $400\,\mathrm{kHz}$.
}

\subsection{MC Samples}
In this study, we utilize the Monte Carlo (MC) method with the KKMC and Tauola generators~\cite{KKMC,Tauola} to {generate $5,567,300$ $e^+e^-\rightarrow\tau^+\tau^-$ events at a CME of $4.68\,\mathrm{GeV}$ based on the theoretical reaction cross-sections and branching ratios, and simulate the detector's electronic signals (including timing, amplitude, etc.) under a fast simulation package~\cite{Shi:2020nrf}. The process $e^+e^-\rightarrow\tau^+\tau^-\left(\tau^+\rightarrow\pi^+\pi^0\bar{\nu}_\tau,\tau^-\rightarrow\pi^-\pi^0\nu_\tau\right)$ accounts for approximately $(25\%)^2=6.2\%$ (i.e., the signal fraction). The main background decays include $\tau^{\pm} \xrightarrow{17.8\%} \nu_{\tau} \mathrm{e}^{\pm} {\nu}_{\mathrm{e}}, \tau^{\pm} \xrightarrow{17.4\%} \nu_{\tau} \mu^{\pm} {\nu}_{\mu}, \tau^\pm\xrightarrow{10.8\%}\pi^\pm \nu_\tau, \tau^\pm\xrightarrow{9.3\%}\pi^\pm \pi^0\pi^0 \nu_\tau$~\cite{PDG}.} The reconstruction software is then used to derive physical quantities such as the momentum and energy of the final state particles from these electronic signals.

The MC simulated data includes truth information (such as particle species) that is not available in actual experiments. This additional information helps optimize the parameter selection in the analysis algorithms, which in turn enhances the effectiveness of these algorithms when applied to real experimental data.

\subsection{Charged Track Selection}
In the final state of the studied process, the detectable particles are $\pi^+$, $\pi^-$, and $\gamma$. The analysis algorithm filters events by reconstructing tracks in the main drift chamber (MDC), selecting those with exactly two charged tracks (corresponding to $\pi^+$ and $\pi^-$) and a total charge sum of zero. It is important to note that only tracks within a small distance from the electron-positron collision point are considered valid to eliminate cosmic rays and beam-related backgrounds. Further, the analysis algorithm uses Particle Identification (PID) based on ionization energy loss ($\mathrm{d}E/\mathrm{d}x$) and Time of Flight (TOF) to identify $\pi^+$ and $\pi^-$ particles. Events are then selected if they contain exactly one $\pi^+$ and one $\pi^-$. {This step can efficiently suppress background events where the final state contains $e$ or $\mu$.}

\subsection{Photon Selection}
For each photon, the algorithm scans all charged tracks to find the minimum angle between the photon and the tracks. A photon is considered valid only if this minimum angle exceeds $20\degree$ and its energy is greater than $0.04\,\mathrm{GeV}$. This criterion helps to reject noise photons produced by hadronic showers. The expected number of final state photons in signal events is $4$ from two $\pi^0$s. Considering the presence of noise photons in the electromagnetic calorimeter (EMC), the algorithm selects events with at least $4$ photons. {This will suppress events without $\pi^0$ in the $\tau$ decay.}

\begin{figure}[!bp]
  \centering
  \includegraphics[width=0.9\textwidth]{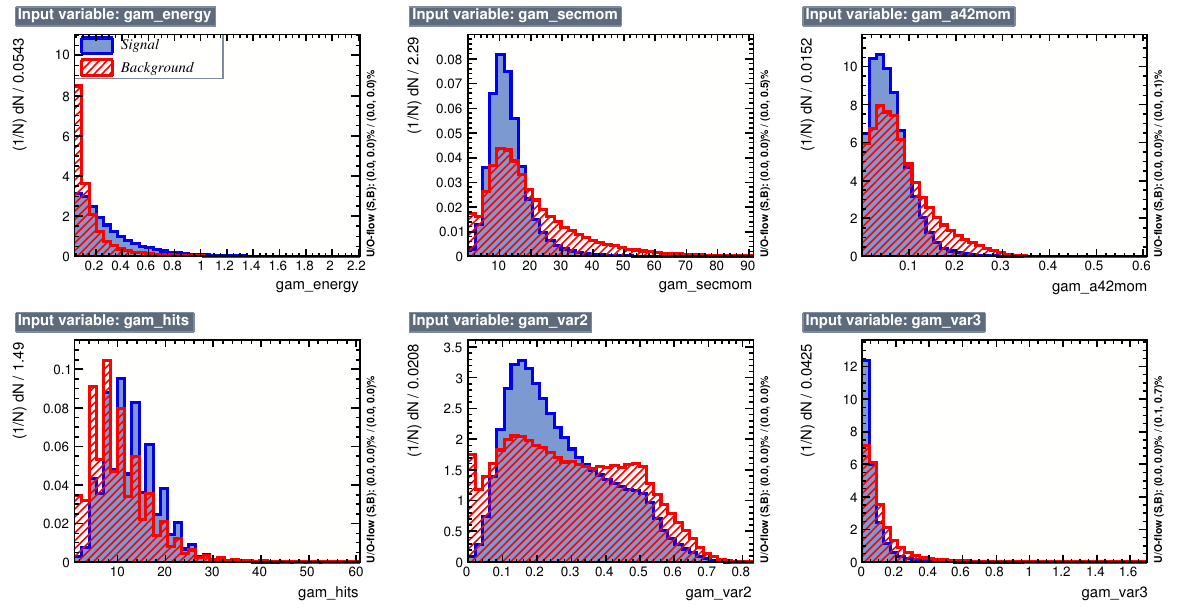}
  \caption{Variable distributions for signal and noise photons.}
  \label{fig:gamma_variable}
\end{figure}

\begin{figure}[!tbp]
  \centering
  \includegraphics[width=0.5\textwidth]{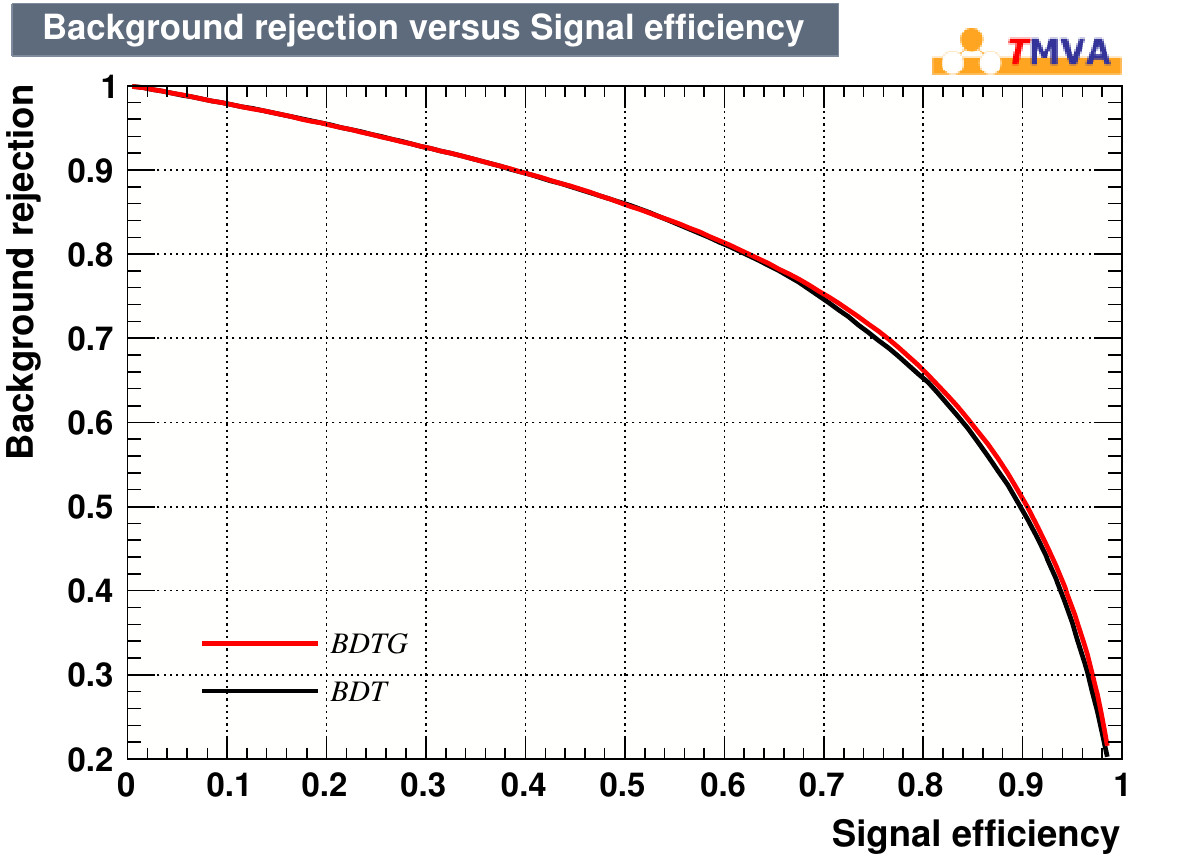}
  \caption{ROC curves for BDT and BDTG for photon selection.}
  \label{fig:gamma_roc}
\end{figure}

{To further eliminate noise photons, a machine learning approach is employed to distinguish signal photons from noise photons. The variables used in the machine learning analysis include:
\begin{itemize}[itemsep=-6pt]
  \item \textbf{gam\_energy}: Energy of the EMC photon cluster (GeV).
  \item \textbf{gam\_secmom}: Second moment of the EMC cluster, describing the shower shape (energy spread).
  \item \textbf{gam\_a42mom}: $a_{42}$ moment of the EMC cluster, a higher-order shower shape variable.
  \item \textbf{gam\_hits}: Number of crystals hit in the cluster.
  \item \textbf{gam\_var2}: $(E_{3\times3} - E_{\text{seed}}) / E_{3\times3}$, energy concentration in $3\times3$ array.
  \item \textbf{gam\_var3}: $(E_{\text{total}} - E_{\text{seed}}) / (N_{\text{hits}} - 1) / E_{\text{seed}}$, average non-seed energy normalized by seed energy.
\end{itemize}
Their distributions for signal and noise photons are shown in Fig.~\ref{fig:gamma_variable}.} The distinct differences between signal and noise photons allow for effective photon selection using machine learning. We train models using Boosted Decision Trees (BDT) and Boosted Decision Trees with Gradient Boosting (BDTG) with the TMVA Toolkit~\cite{TMVA}, and the resulting Receiver Operating Characteristic (ROC) curves are shown in Fig.~\ref{fig:gamma_roc}. BDTG performs slightly better than BDT, so BDTG is used in subsequent analysis.

\begin{figure}[!bp]
  \centering
  \includegraphics[width=0.48\textwidth]{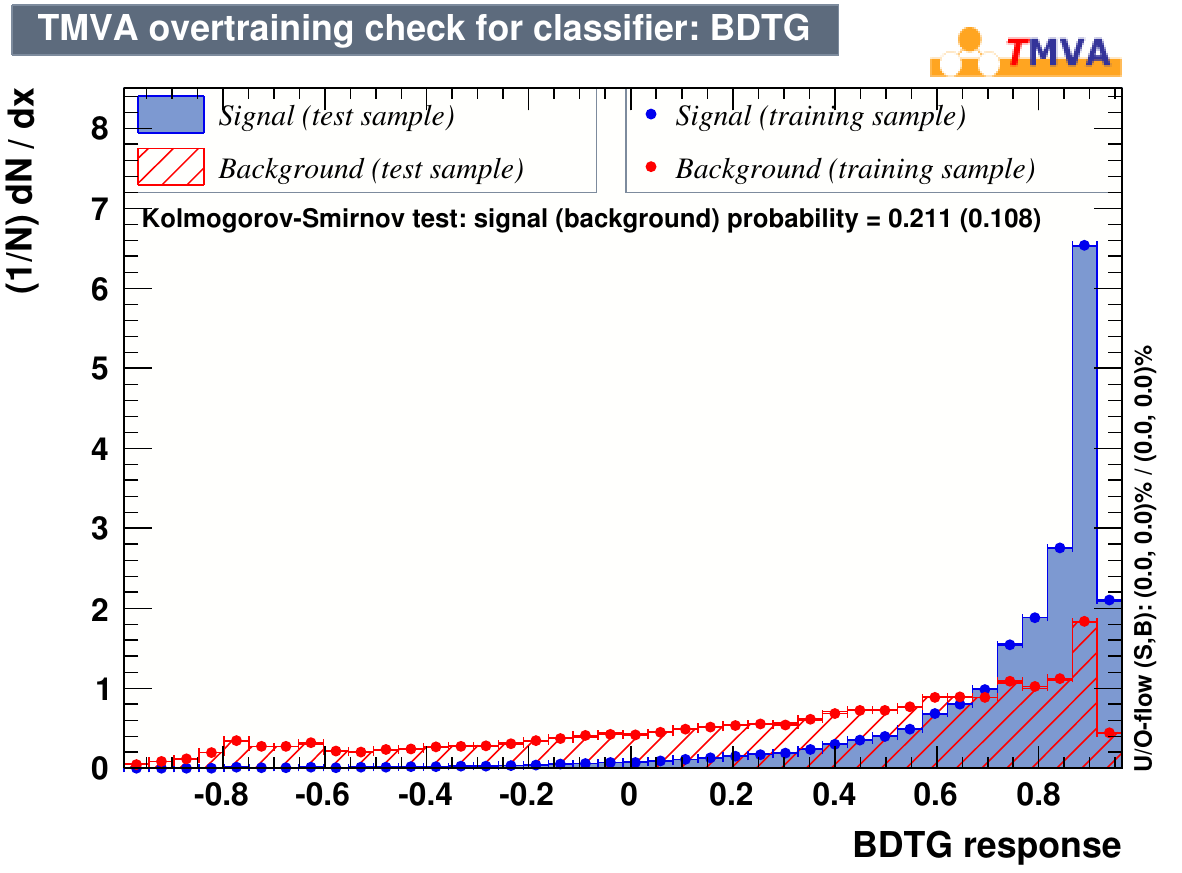}
  \includegraphics[width=0.48\textwidth]{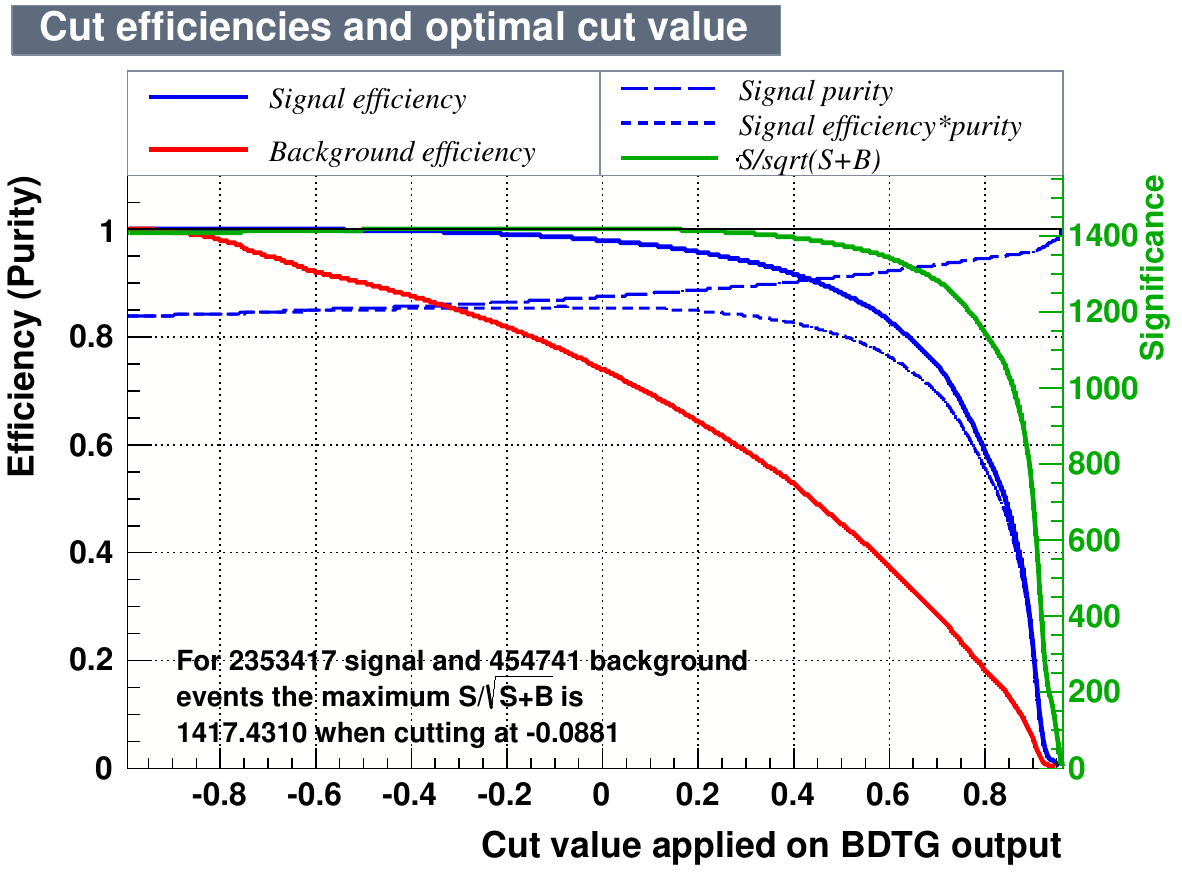}
  \caption{BDTG response (left panel) and cut efficiency (right panel) for photon selection.}
  \label{fig:gamma_BDTG_response_cut}
\end{figure}

The BDTG response distributions for training and testing datasets are shown in the left panel of Fig.~\ref{fig:gamma_BDTG_response_cut}, indicating no significant overfitting as the distributions agree quite well. The BDTG cut efficiency is presented in the right panel of Fig.~\ref{fig:gamma_BDTG_response_cut}. Considering both signal efficiency and signal-to-noise ratio, a BDTG cut of $0.6$ is chosen, leaving approximately $84\%$ of signal photons and $37\%$ of noise photons.

After the machine learning selection, the effective number of photons for $\rho\rho$ mode events is typically $4$. Consequently, the analysis algorithm selects events with $4$ photons.

\subsection{Particle Pairing}
While we can reconstruct the tracks of final-state $\pi^+$, $\pi^-$, and photons, it is unclear which photons are signal photons and how these photons associate with the corresponding $\pi^+$ or $\pi^-$ in $\rho\rho$ mode. The objective of this step is to determine the best pairing of these particles. Various methods were designed and compared for selecting the optimal pairing, with the best approach determined by the $\chi^2$ value from a joint kinematic fitting, as detailed in Section \ref{sec:pairing}. The joint kinematic fitting not only determines the best pairing but also improves the signal-to-noise ratio by selecting events with $\chi^2<10$.

\subsection{Event-Level Machine Learning Selection}
\begin{figure}[htbp]
  \centering
  \includegraphics[width=0.48\textwidth]{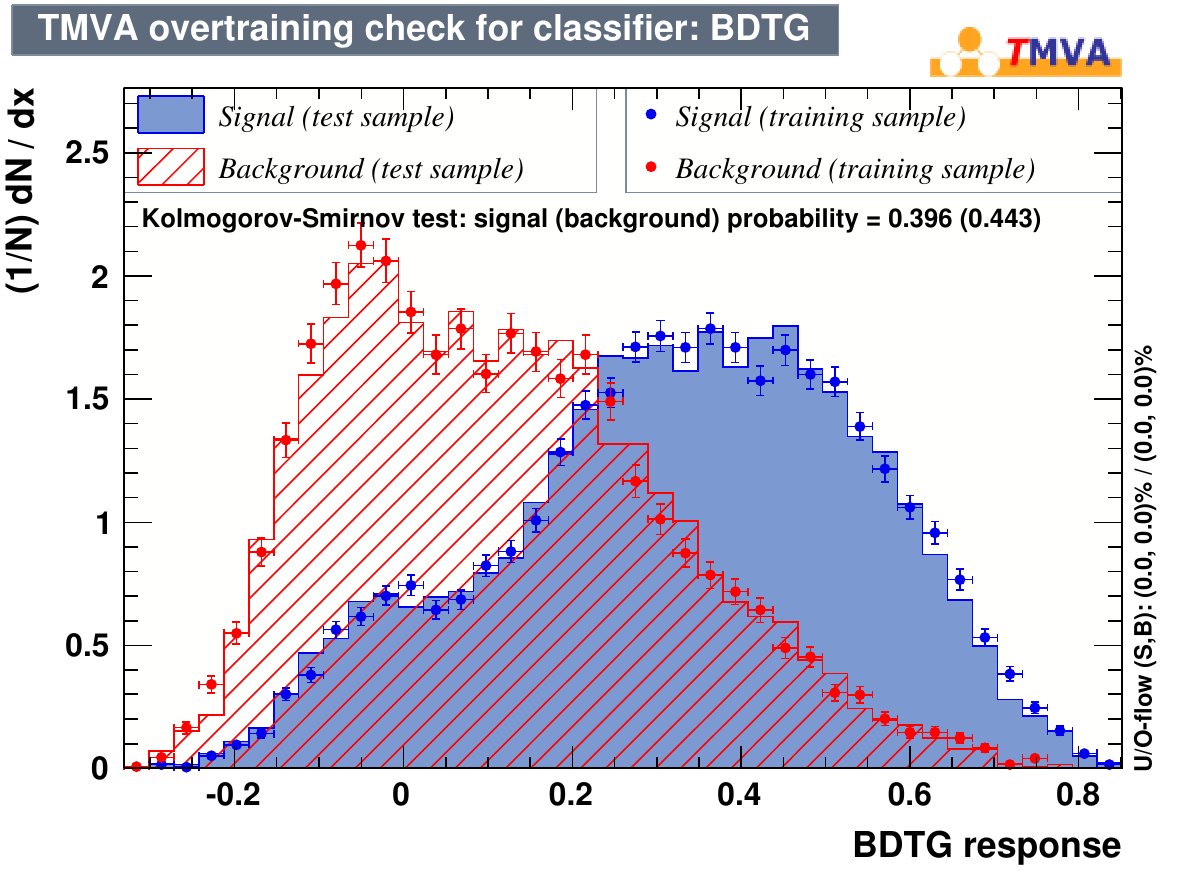}
  \includegraphics[width=0.48\textwidth]{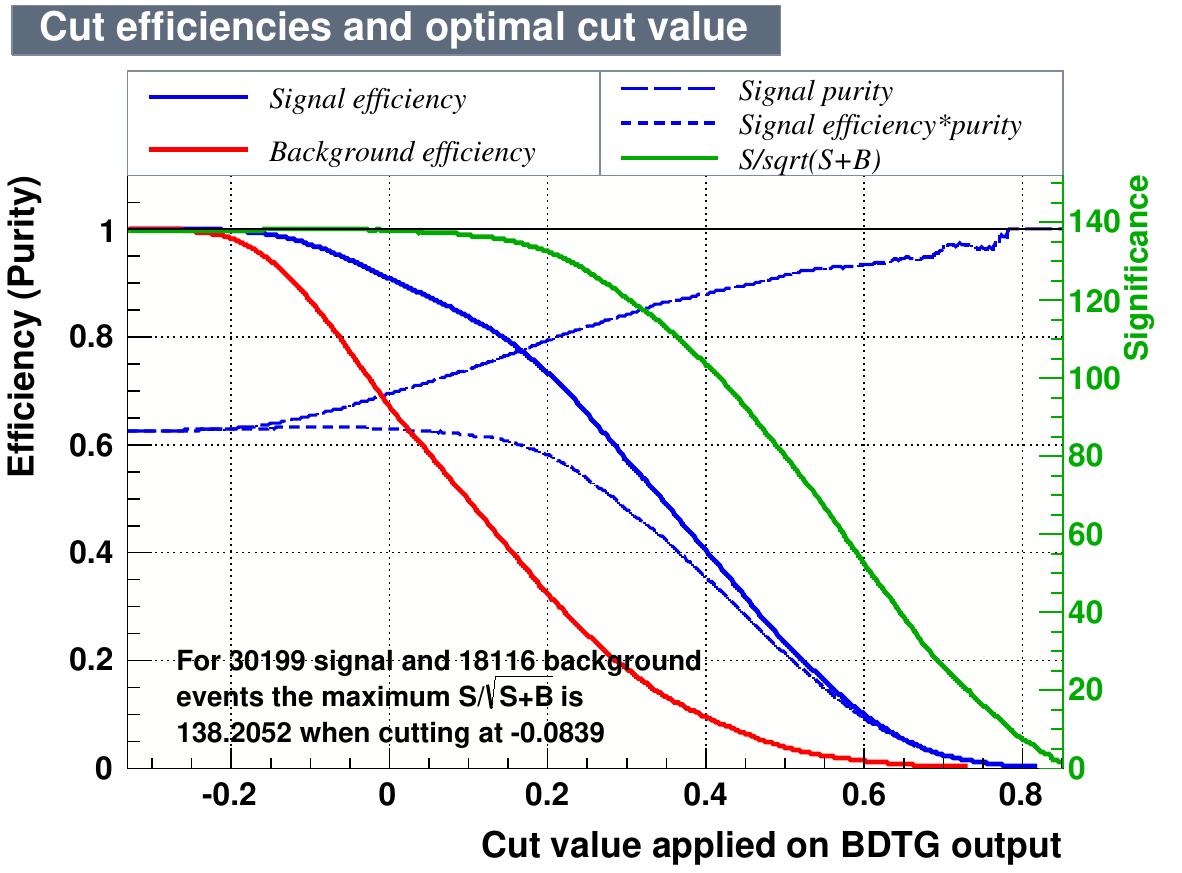}
  \caption{BDTG response (left panel) and cut efficiency (right panel) for event selection.}
  \label{fig:event_BDTG_response_cut}
\end{figure}

To further enhance the signal-to-noise ratio, event-level machine learning is applied to select signal events. The variables used in this analysis include the momentum of the $\pi^+$, $\pi^-$, and the four photons (in sequence). Similar to the photon selection using machine learning, the BDTG response and cut efficiency are shown in Fig.~\ref{fig:event_BDTG_response_cut}. A BDTG cut of $0.2$ is selected, resulting in about $73\%$ of the signal events and $32\%$ of the background events remaining.

\subsection{$\tau$ Momentum Reconstruction}
To obtain the optimal observables, it is necessary to reconstruct for the momentum of the $\tau$ lepton, as detailed in Section \ref{sec:tau_momentum}. During this process, scenarios may arise with two solutions, one solution, or no solution. Events with no solution are considered background and are excluded.

\subsection{Event Selection Results}
The selection algorithms applied reduced the initial $5,567,300$ events to $27,125$ events. The selection efficiency at each step is summarized in Table \ref{tab:event_selection}. The overall selection efficiency is $0.49\%$, and the signal ($\tau^+\rightarrow\pi^+\pi^0\bar{\nu}_\tau,\tau^-\rightarrow\pi^-\pi^0\nu_\tau$) efficiency is $6.3\%$, with the signal purity increased from $6.2\%$ to $80.0\%$.

\begin{table}[hbtp]
  \centering
  \caption{Event selection results for $5,567,300$ simulated events.}
  \begin{small}
    \begin{tabular}{|c|c|c|c|c|}
      \hline
      \multirow{2}{*}{No.} & \multirow{2}{*}{Step} & \multicolumn{2}{c|}{Percentage of previous step} & \multirow{2}{*}{Signal purity} \\ \cline{3-4} 
  
                 &                       & Inclusive events & Signal events & \\ \hline
  
      0   &         Total events         &   -    &  -  & \textbf{6.2\%} \\ \hline
  
      1   & Number of charged tracks = 2, total charge = 0 &   58.3\% & 76.5\% & 8.1\% \\ \hline
  
      2   &         Number of photons = 4         &    7.2\%  & 23.7\% & 26.7\% \\ \hline
  
      3   &   Number of $\pi^+$ = 1, Number of $\pi^-$ = 1   &    81.8\% & 92.1\% & 30.0\% \\ \hline
  
      4   &    Passed the particle pairing    &    25.2\% & 52.3\% & 62.5\% \\ \hline
  
      5   &  Passed event-level machine learning selection  &    57.9\% & 73.4\% & 79.3\% \\ \hline
  
      6   &  Passed the $\tau$ momentum reconstruction  &    97.0\% & 97.7\% & \textbf{80.0\%} \\ \hline
    \end{tabular}
  \end{small}
  \label{tab:event_selection}
\end{table}

{Detailed event type analysis on the $27,125$ selected events with a generic topology analysis package, TopoAna~\cite{TopoAna}, shows that background decays $\tau^{\pm} \rightarrow \nu_{\tau} \mathrm{e}^{\pm} {\nu}_{\mathrm{e}}, \tau^\pm\rightarrow\pi^\pm \nu_\tau$ have been almost entirely filtered out. The dominant background processes after selection involves more $\pi^0$, such as $\tau^\pm\rightarrow\pi^\pm \pi^0\pi^0 \nu_\tau$ (about $14\%$), which may require further selection optimization.}

Considering that at the CME of $4.2\,\mathrm{GeV}$, the STCF is expected to produce approximately $3.5 \times 10^9$ $\tau$ lepton pairs per year~\cite{STCF}, of which around $2.2 \times 10^8$ are signal events of the $\rho\rho$ mode, we estimate a signal yield of $1.4 \times 10^7$ per year after selection, with a signal efficiency of $6.3\%$ and a signal purity of $80.0\%$. For comparison, up to 2022, Belle experiments have $5.2 \times 10^7$ events of the $\rho\rho$ mode. After selection, the signal yield is $3.3 \times 10^6$, with a signal efficiency of $6.3\%$ and a signal purity of $82.4\%$~\cite{Belle2022}. The current feasibility study shows that the tau pair selection efficiency is comparable to that of the B factory. After 10 years of operation, the STCF will collect $1.4 \times 10^8$ tau pairs of $\rho\rho$ mode after reconstruction, which is $2$ orders of magnitude higher than Belle. In addition, the event selection at STCF can be further optimized with the vertex detector. This increase in statistics will improve the sensitivity of the $\tau$ lepton EDM measurement.

\section{Particle Pairing}
\label{sec:pairing}
This section introduces and compares four methods for determining the pairings of $\pi^+, \gamma_{(1)}, \gamma_{(2)}$ and $\pi^-, \gamma_{(3)}, \gamma_{(4)}$. {The main method we use is joint kinematic fitting.}

Let the number of photons detected in the electromagnetic calorimeter (EMC) for the current event be $n$ ($n \geq 4$, and $n=4$ in this study). The total number of possible pairings is $C_n^2 C_{n-2}^2$. The analysis algorithm performs kinematic fitting on all possible pairings. The kinematic constraints are imposed based on the conservation laws, including the total energy-momentum conservation and the mass constraints of the intermediate states $\tau$ and $\pi^0$. The specific equations are as follows:
\begin{equation}
  \begin{aligned}
    \boldsymbol{p}_{\pi^+}+\boldsymbol{p}_{\pi^-}+\boldsymbol{p}_{\pi^0_{(1)}}+\boldsymbol{p}_{\pi^0_{(2)}}+\boldsymbol{p}_{\nu_{(1)}}+\boldsymbol{p}_{\nu_{(2)}}&=\boldsymbol{p}^\text{Total},\\
    E_{\pi^+}+E_{\pi^-}+E_{\pi^0_{(1)}}+E_{\pi^0_{(2)}}+E_{\nu_{(1)}}+E_{\nu_{(2)}}&=E^\text{Total},
  \end{aligned}
  \label{pETotal}
\end{equation}
\begin{equation}
  E_{\nu_{(1,2)}}^2=\boldsymbol{p}_{\nu_{(1,2)}}^2 c^2,
  \label{nu}
\end{equation}
\begin{equation}
  E_{\gamma_{(1,2,3,4)}}^2=\boldsymbol{p}_{\gamma_{(1,2,3,4)}}^2 c^2,
  \label{gamma}
\end{equation}
\begin{equation}
  E_{\pi^\pm}^2 = \boldsymbol{p}_{\pi^\pm}^2 c^2 + m_{\pi}^2 c^4,
  \label{pipm}
\end{equation}
\begin{equation}
  \begin{aligned}
    E_{\pi^0_{(1)}}^2 &= \boldsymbol{p}_{\pi^0_{(1)}}^2 c^2 + m_{\pi}^2 c^4,\\
    E_{\pi^0_{(2)}}^2 &= \boldsymbol{p}_{\pi^0_{(2)}}^2 c^2 + m_{\pi}^2 c^4,
  \end{aligned}
  \label{pi0}
\end{equation}
\begin{equation}
  \begin{aligned}
    \left(E_{\pi^+}+E_{\pi^0_{(1)}}+E_{\nu_{(1)}}\right)^2 &= \left(\boldsymbol{p}_{\pi^+}+\boldsymbol{p}_{\pi^0_{(1)}}+\boldsymbol{p}_{\nu_{(1)}}\right)^2 c^2 + m_{\tau}^2 c^4,\\
    \left(E_{\pi^-}+E_{\pi^0_{(2)}}+E_{\nu_{(2)}}\right)^2 &= \left(\boldsymbol{p}_{\pi^-}+\boldsymbol{p}_{\pi^0_{(2)}}+\boldsymbol{p}_{\nu_{(2)}}\right)^2 c^2 + m_{\tau}^2 c^4,
  \end{aligned}
  \label{tau}
\end{equation}
where:
\begin{equation}
  \begin{aligned}
    &\boldsymbol{p}_{\pi^0_{(1)}}=\boldsymbol{p}_{\gamma_{(1)}}+\boldsymbol{p}_{\gamma_{(2)}},\quad E_{\pi^0_{(1)}}=E_{\gamma_{(1)}}+E_{\gamma_{(2)}},\\
    &\boldsymbol{p}_{\pi^0_{(2)}}=\boldsymbol{p}_{\gamma_{(3)}}+\boldsymbol{p}_{\gamma_{(4)}},\quad E_{\pi^0_{(2)}}=E_{\gamma_{(3)}}+E_{\gamma_{(4)}}.
  \end{aligned}
\end{equation}

Eqs. (\ref{gamma}) and (\ref{pipm}) are automatically satisfied by the reconstructed data, while Eq.~(\ref{pi0}) imposes the $\pi^0$ mass constraint on the photon energy-momentum. Eqs. (\ref{pETotal}), (\ref{nu}), and (\ref{tau}) constrain the unknown neutrino ($\nu$) energy-momentum. The kinematic fitting adjusts the known quantities, solves for the unknowns, and yields a fitting error $\chi^2$ (the deviation of fitted and original values over the error). The correct pairing will have a smaller $\chi^2$ and a smaller $p_{\nu}=p_{\nu_{(1)}} + p_{\nu_{(2)}}$, whereas incorrect pairings lead to situations where the constraint equations have no solutions or yield non-physical solutions, resulting in a larger $\chi^2$ and a larger $p_{\nu}$.

From the truth information in the simulated data, it is observed that the sum of the magnitudes of the momenta of the two neutrinos is generally less than $E^\text{Total}/2$. Therefore, the kinematic fitting solution is required to satisfy:
\begin{equation}
  p_{\nu}=p_{\nu_{(1)}}+p_{\nu_{(2)}}<\frac{E^\text{Total}}{2}.
  \label{pnu_cut}
\end{equation}

Under this constraint, the algorithm selects the pairing with the smallest $\chi^2$ as the chosen pairing. The kinematic fitting can also be used to reject events that fail the fit or have $\chi^2 \geq 10$ or $p_{\nu}\geq E^\text{Total}/2$, thus improving the signal-to-noise ratio.

\begin{table}[bp]
  \centering
  \caption{Kinematic fitting pass rate and pairing correct rate.}
  \begin{tabular}{|c|c|}
    \hline
    Signal events & Percentage \\ \hline
    Passed the joint kinematic fitting    & 52.3\%   \\ \hline
    Correct $\pi^0$ pairing     & 95.6\%  \\ \hline
    Fully correct particle pairing     & 82.5\%  \\ \hline
  \end{tabular}
  \label{tab:kfit}
\end{table}

The truth information in the simulated data contains the actual pairing information of photons with $\pi^+, \pi^-$. By comparing selected pairings with this information (pairings with relative and absolute errors of each photon momentum component in order within an acceptable range are considered correct), the pairing correct rate can be obtained. Additionally, the correctness of the photon pairing forming the $\pi^0$ can be examined, that is, only checking whether $\gamma_{(1)}, \gamma_{(2)}$ are paired to form a $\pi^0_{(1)}$ and whether $\gamma_{(3)}, \gamma_{(4)}$ are paired to form a $\pi^0_{(2)}$, regardless of the pairing selection of $\pi^0_{(1)}, \pi^0_{(2)}$ with $\pi^+, \pi^-$. For signal events, the pass rate and pairing correct rate of the kinematic fitting are shown in Table \ref{tab:kfit}, with a fully correct particle pairing rate of up to 82.5\%.

\begin{figure}[tbp]
    \centering
    \includegraphics[width=0.49\linewidth]{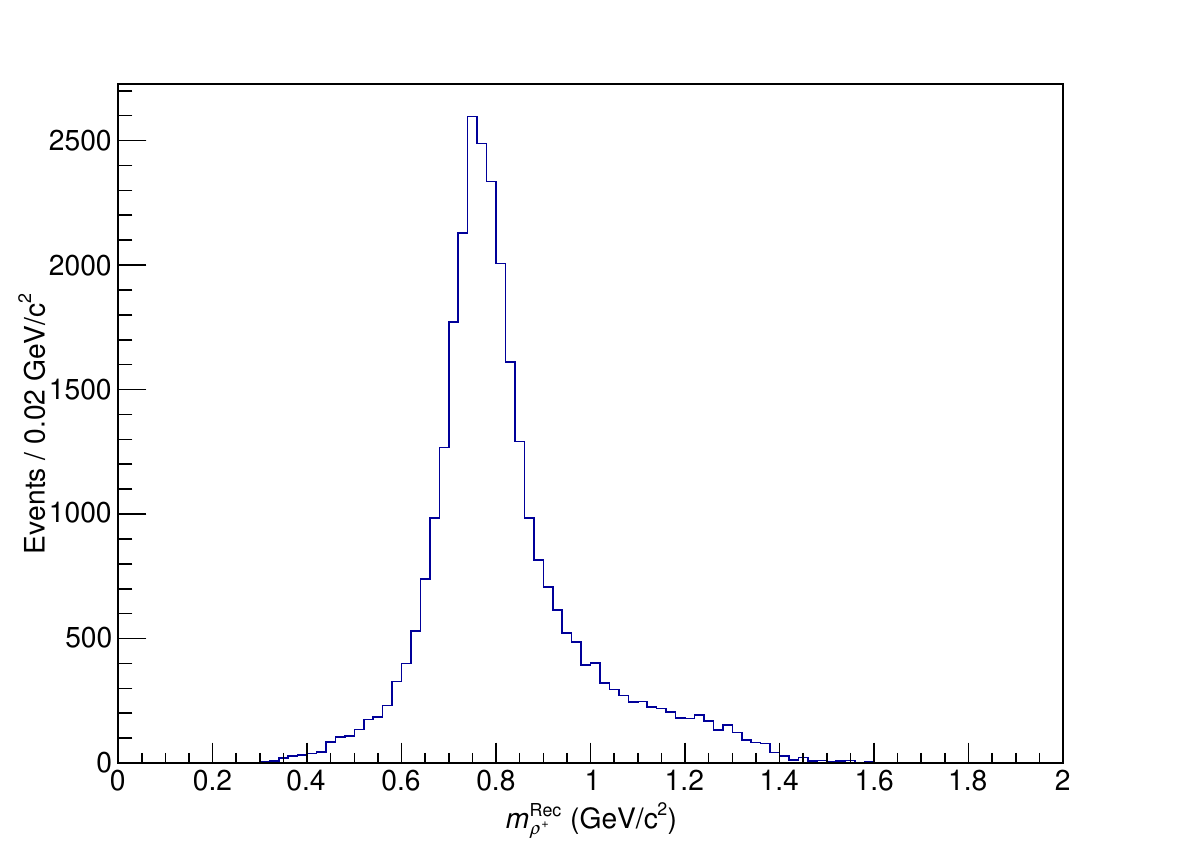}
    \includegraphics[width=0.49\linewidth]{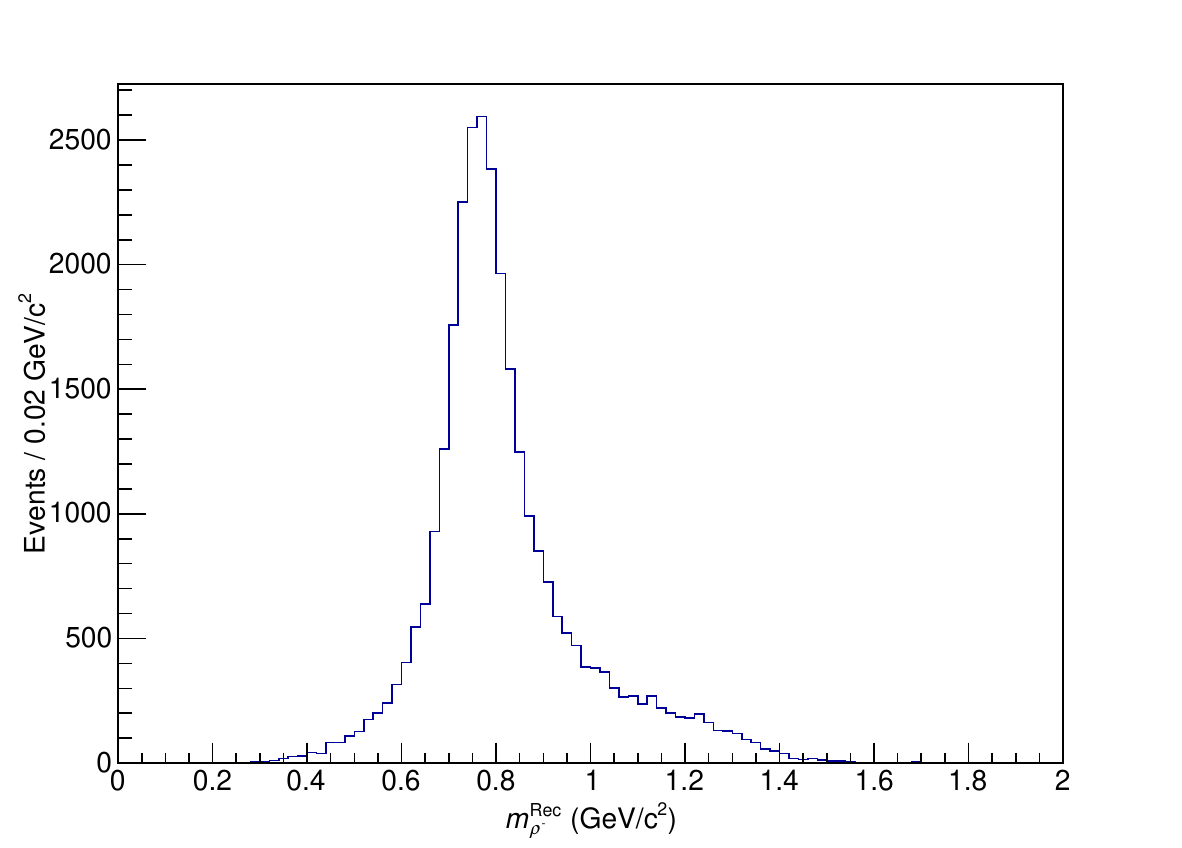}
    \caption{Invariant mass distribution of $(\pi^+,\pi^0_{(1)})$ (left panel) and $(\pi^-,\pi^0_{(2)})$ (right panel).}
    \label{fig:mrho}
\end{figure}

\begin{figure}[tbp]
    \centering
    \includegraphics[width=0.49\linewidth]{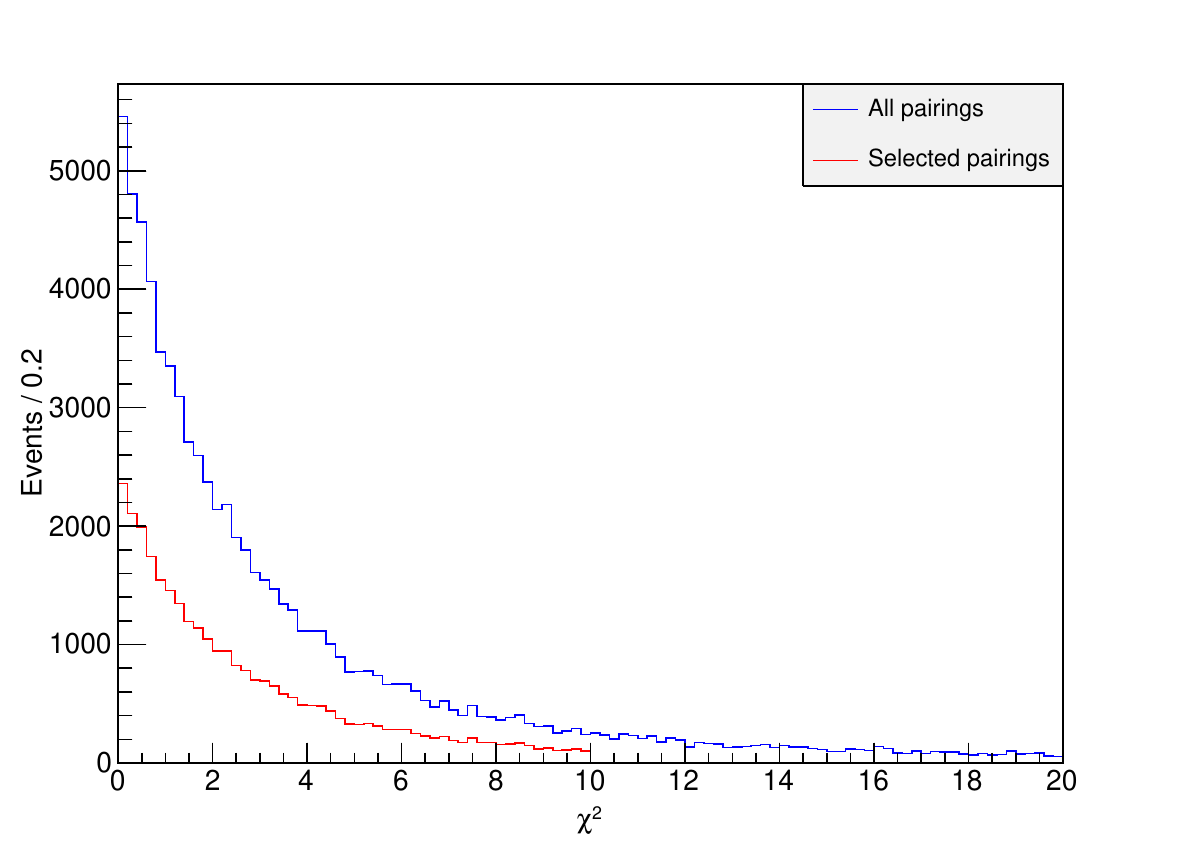}
    \includegraphics[width=0.49\linewidth]{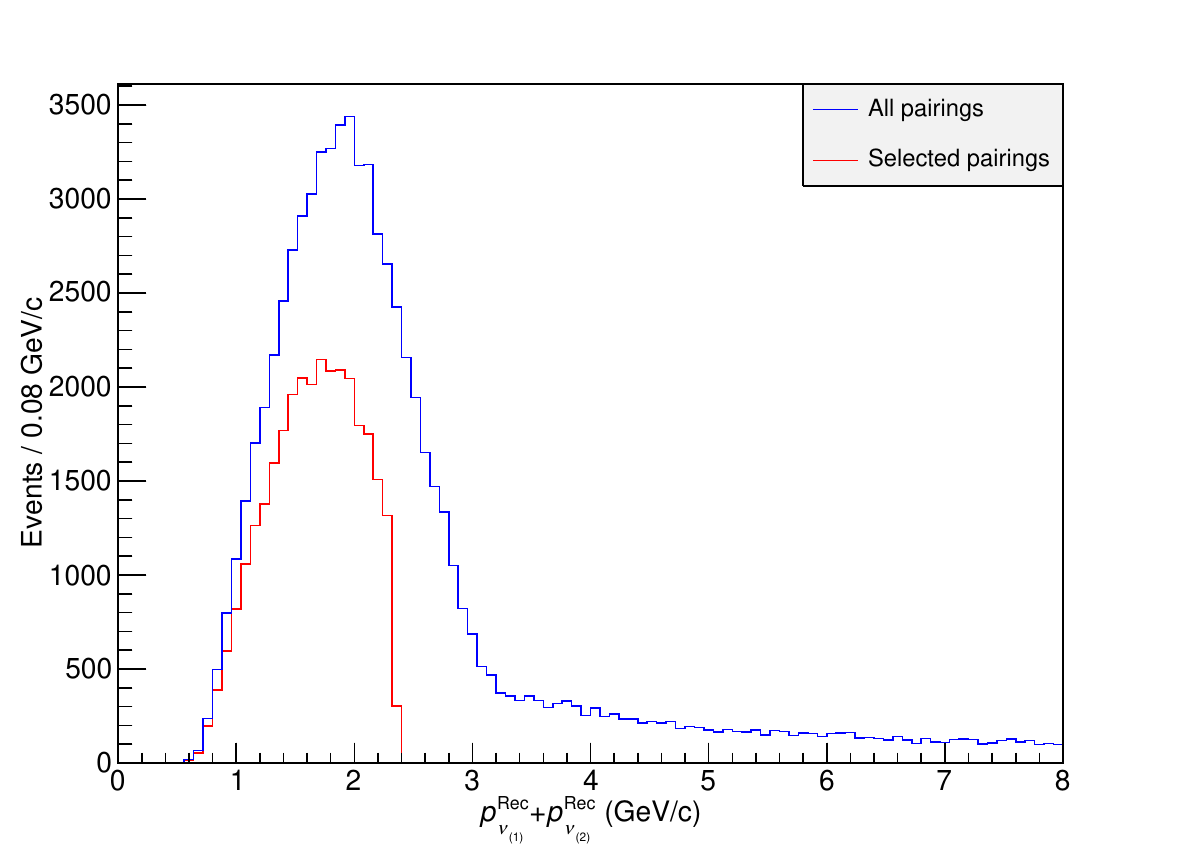}
    \caption{Distribution of $\chi^2$ (left panel) and $p_{\nu}=p_{\nu_{(1)}} + p_{\nu_{(2)}}$ (right panel) for selected pairings and all pairings.}
    \label{fig:pnu}
\end{figure}

We examine the invariant mass distributions of $(\pi^+, \pi^0_{(1)})$ and $(\pi^-, \pi^0_{(2)})$, as shown in Fig.~\ref{fig:mrho}. It can be seen that the invariant masses are concentrated around the mass of the $\rho(770)$, as expected. Besides, we compare the distribution of $\chi^2$ and $p_{\nu}=p_{\nu_{(1)}} + p_{\nu_{(2)}}$ for selected pairings and all pairings in Fig.~\ref{fig:pnu}, which also meet expectations.

{We tested three additional methods for particle pairing (detailed in Appendix~\ref{app:pairing}): stepwise kinematic fitting, kinematic fitting with $\rho$ resonance mass constraint, and kinematic fitting with momentum direction. Among the methods tested, joint kinematic fitting achieves the highest accuracy. Therefore, all subsequent steps are based on the data selected using this method.}

\section{The Measurement of $\tau$ EDM}
\subsection{$\tau$ Momentum Reconstruction}
\label{sec:tau_momentum}
Due to the presence of neutrinos among the $\tau$ lepton decay products, which cannot be detected by the instruments, it is impossible to fully reconstruct the final-state particles to determine the momentum of the $\tau$ lepton. Therefore, a special method is required to calculate the $\tau$ lepton momentum.

After event selection and particle pairing, the momenta and energies of $\pi^+, \pi^-, \pi^0_{(1)}$, and $\pi^0_{(2)}$ have been obtained from the track information of the main drift chamber and the photon signals in the electromagnetic calorimeter. To accurately determine the $\tau$ lepton momentum, analytical computation is used to derive the solutions.

In electron-positron collider experiments, the beams are designed to collide at a small angle at the interaction point to optimize collision performance and data collection efficiency. In the laboratory frame, the total momentum of the system is given by $\boldsymbol{p}^\text{Total} = (p_x, 0, 0)$, where $p_x$ is a small value that can be determined from the total energy $E^\text{Total}$ and the design parameters of the experiment. By applying a Lorentz transformation to all particles to move into the center-of-mass frame, we have:
\begin{equation}
  \boldsymbol{p}^\text{Total}=\boldsymbol{0},
  \label{p=0}
\end{equation}
and the energies of the $\tau^+$ and $\tau^-$ leptons become:
\begin{equation}
  E_{\tau^+}=E_{\tau^-}=E_{\tau}=\frac{E^\text{Total}}{2}.
  \label{Etau}
\end{equation}

The magnitude of the $\boldsymbol{p}_{\tau^\pm}$ can then be derived from $E_{\tau}$ as follows:
\begin{equation}
  \boldsymbol{p}_{\tau^\pm}^2 c^2=E_{\tau}^2-m_{\tau}^2 c^4.
\end{equation}

Next, we solve for the direction of $\boldsymbol{p}_{\tau^\pm}$. By combining Eqs.~(\ref{pETotal}), (\ref{nu}), (\ref{tau}), and (\ref{p=0}), the cosine of the angle $\theta_{\pm}$ between the vectors $\boldsymbol{p}_{\tau^\pm}$ and $\boldsymbol{p}_{h^\pm} = \boldsymbol{p}_{\pi^\pm} + \boldsymbol{p}_{\pi^0}$ is given by~\cite{ImpactParameters}:
\begin{equation}
  \cos \theta_{\pm}=\frac{\gamma x_{\pm}-\left(1+r_{\pm}^2\right) / 2 \gamma}{\beta \sqrt{\gamma^2 x_{\pm}^2-r_{\pm}^2}},
\end{equation}
where:
\begin{equation}
  x_{\pm}=\frac{E_{h^\pm}}{E_\tau}=\frac{E_{\pi^\pm}+E_{\pi^0}}{E_\tau},\quad r_{\pm}=\frac{m_{\pi}}{m_\tau},
\end{equation}
\begin{equation}
  \gamma = \frac{E_{\tau}}{m_{\tau}},\quad \beta = \sqrt{1 - \frac{1}{\gamma^2}}.
\end{equation}

In the center-of-mass frame, we have:
\begin{equation}
  \boldsymbol{p}_{\tau^+}=-\boldsymbol{p}_{\tau^-}.
\end{equation}
Given that the angle between the vectors $\boldsymbol{p}_{\tau^+}$ and $\boldsymbol{p}_{h^+}$ is $\theta_+$, and the angle between the vectors $-\boldsymbol{p}_{\tau^-}$ and $-\boldsymbol{p}_{h^-}$ is $\theta_-$, $\boldsymbol{p}_{\tau^+}$ must lie on the intersection of two cones: one with axis $\boldsymbol{p}_{h^+}$ and half-angle $\theta_+$, and the other with axis $-\boldsymbol{p}_{h^-}$ and half-angle $\theta_-$. Using Mathematica, two analytic solutions for $\boldsymbol{p}_{\tau^+}$ were obtained. There are square root operations in the expressions for the solutions. If the term under the square root is negative (small negative values due to experimental uncertainties are treated as zero in this study), the corresponding case has no physical solution and is therefore discarded. If the term is positive, it yields two distinct solutions. Among these, only one corresponds to the physical reality. Experimentally, this ambiguity cannot be resolved, so following the same approach as the Belle experiment~\cite{Belle2022}, this study adopts the method of taking the average of the two solutions.

\begin{figure}[tbp]
    \centering
    \includegraphics[width=0.49\linewidth]{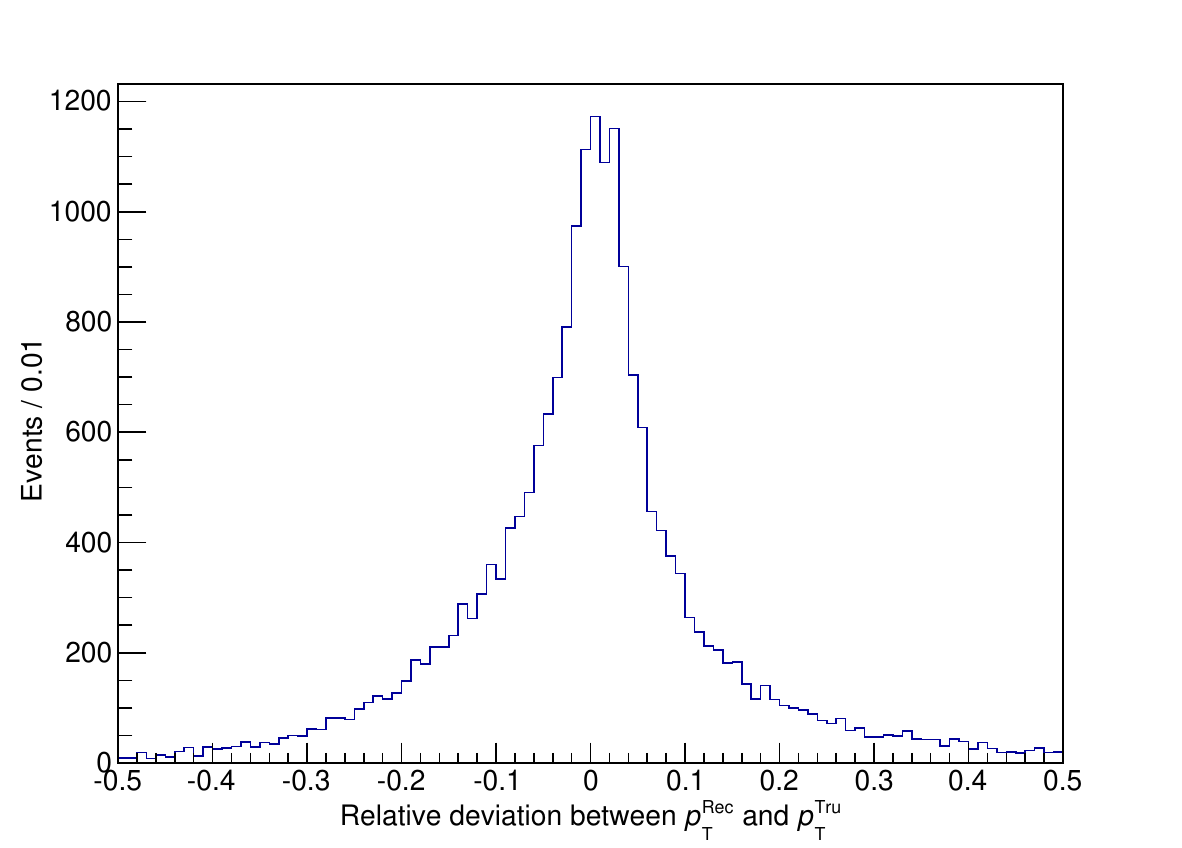}
    \includegraphics[width=0.49\linewidth]{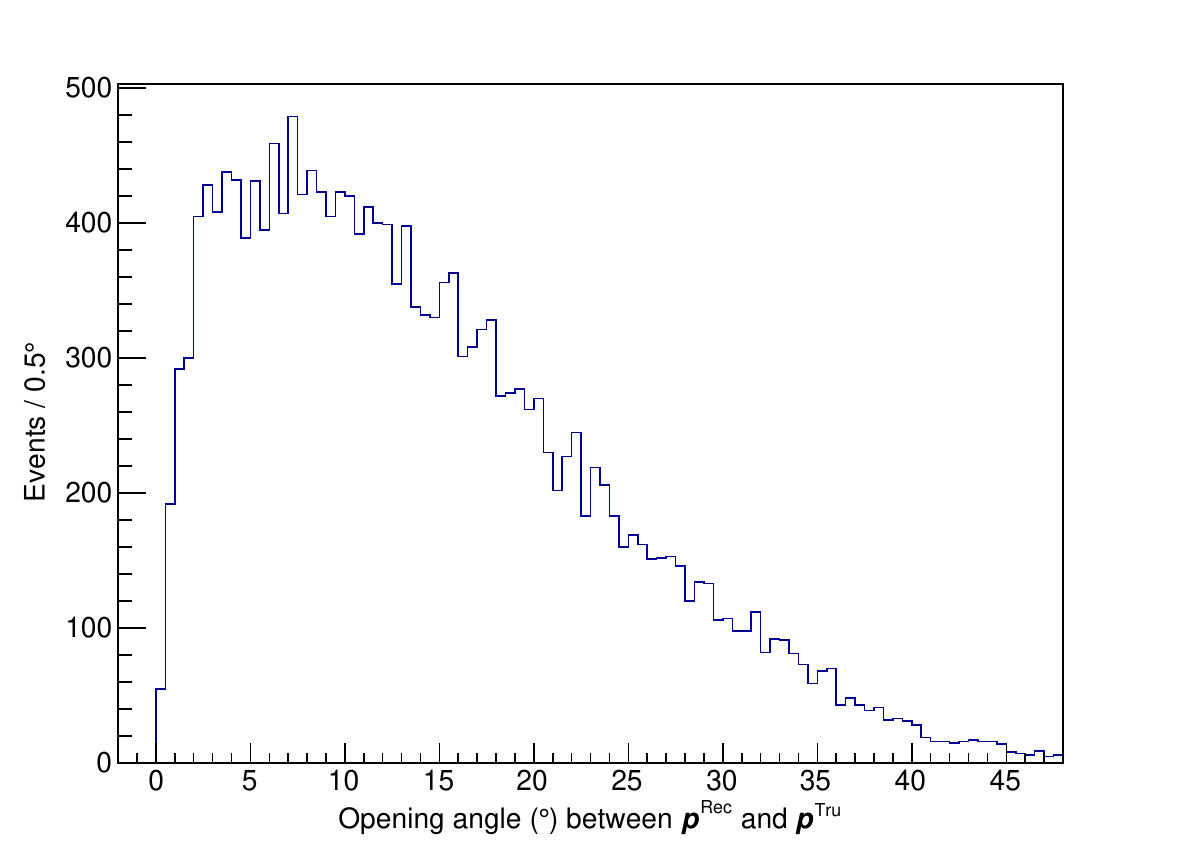}
    \caption{Relative deviation of the transverse momentum (left panel) and the direction (right panel) of the $\tau$ lepton.}
    \label{fig:tau_compare_Truth}
\end{figure}

After determining $\boldsymbol{p}_{\tau^\pm}$ in the center-of-mass frame, a Lorentz transformation is applied to convert it back to the laboratory frame, resulting in the reconstructed momentum of the $\tau$ lepton. For signal events, the relative deviation of the reconstructed transverse momentum $p_T = \sqrt{p_x^2 + p_y^2}$ and the angular distribution between the reconstructed and truth $\tau$ lepton momentum are shown in Fig.~\ref{fig:tau_compare_Truth}. The full width at half maximum (FWHM) of the relative deviation in transverse momentum is $0.10$, and the peak opening angle is $7^\circ$ with a root mean square (RMS) of $10^\circ$.

\subsection{Spin Correlation in the Decay}
The decay products keep the information of the spin of the tau lepton, which can be used to reconstruct the polarimeter vector. For the hadronic decay channel of the tau lepton, we have the decay matrix element as~\cite{Tauola}:
\begin{align}
|\mathcal{M}_\pm|^2 &= G_F^2(\omega_\pm + H_\pm\cdot s_\pm) = G_F^2 \omega_\pm (1+ h_\pm\cdot s_\pm),\\
\omega_\pm &= p_\pm^\mu(\Pi_\mu^\pm - \Pi_\mu^{\pm 5}),\\
H^\mu_\pm &= \pm \frac{m_\tau^2g^{\mu\nu}-p_\pm^\mu p_\pm^\nu}{m_\tau}(\Pi^{\pm 5}_\nu - \Pi^{\pm}_\nu),
\end{align}
with:
\begin{align}
\Pi_\mu^\pm &= 2\left[(J^{\pm*}\cdot N^\pm)J_\mu^\pm + (J^\pm\cdot N^\pm)J_\mu^{\pm*} - (J^{\pm*}\cdot J^\pm)N^\pm_\mu\right],\\
\Pi_\mu^{\pm5} &= 2\Im\left[\epsilon_{\mu\nu\rho\sigma}J^{\pm*\nu}J^{\rho}N^\sigma\right],
\end{align}
where $p_\pm$ are the momenta of $\tau^\pm$, $N^\pm$ are the momenta of the corresponding neutrinos. $J$ will be different for different decay channels:
\begin{align}
    J^\pm_\mu(\tau\to\pi\nu) &\propto p^\pi_\mu,\\
    J^\pm_\mu(\tau^\pm\to\pi^\pm\pi^0\nu) &\propto p^{\pi^\pm}_\mu - p^{\pi^0}_\mu.
\end{align}

Combining the decay and the production, we obtain for the each component of the produciton matrix element:
\begin{align}
\mathcal{M}_i^2 = \overline{\mathcal{M}_i^2}\omega_+\omega_-(1-a_{i\mu}h_+^\mu - b_{i\mu} h_-^\mu + c_{i\mu\nu}h_+^\mu h_-^\nu).
\end{align}
Then the optimal observable can be constructed by all the final state momentum. However, as discussed above, we can only determine the momenta of the neutrino/tau lepton up to a two-fold ambiguity. Then the matrix element calculated from these two solutions will be averaged when calculating the optimal observable.

\subsection{The Relationship Between Optimal Observables and the EDM}
\begin{figure}[htbp]
  \centering
  \includegraphics[width=0.48\linewidth]{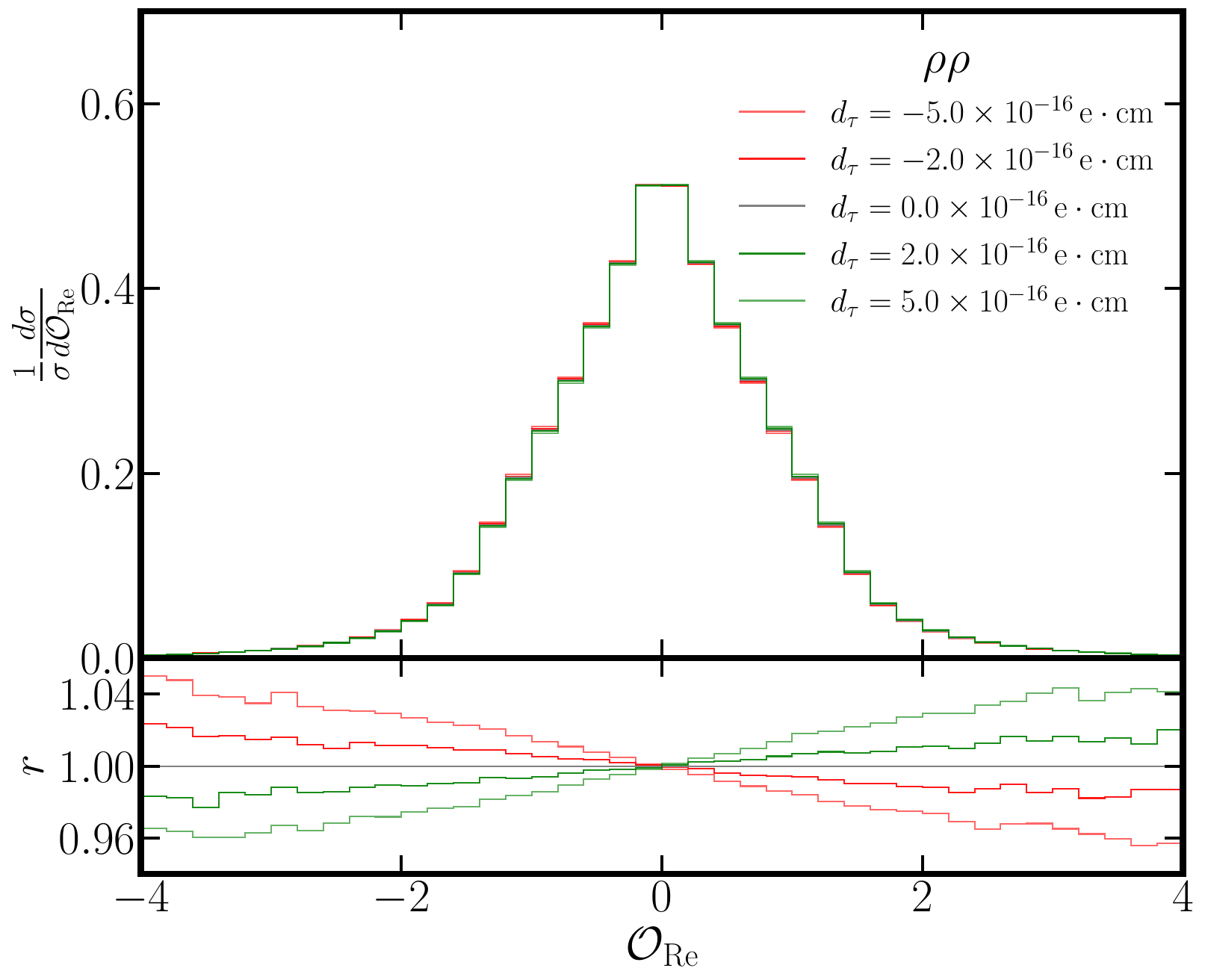}
  \includegraphics[width=0.49\linewidth]{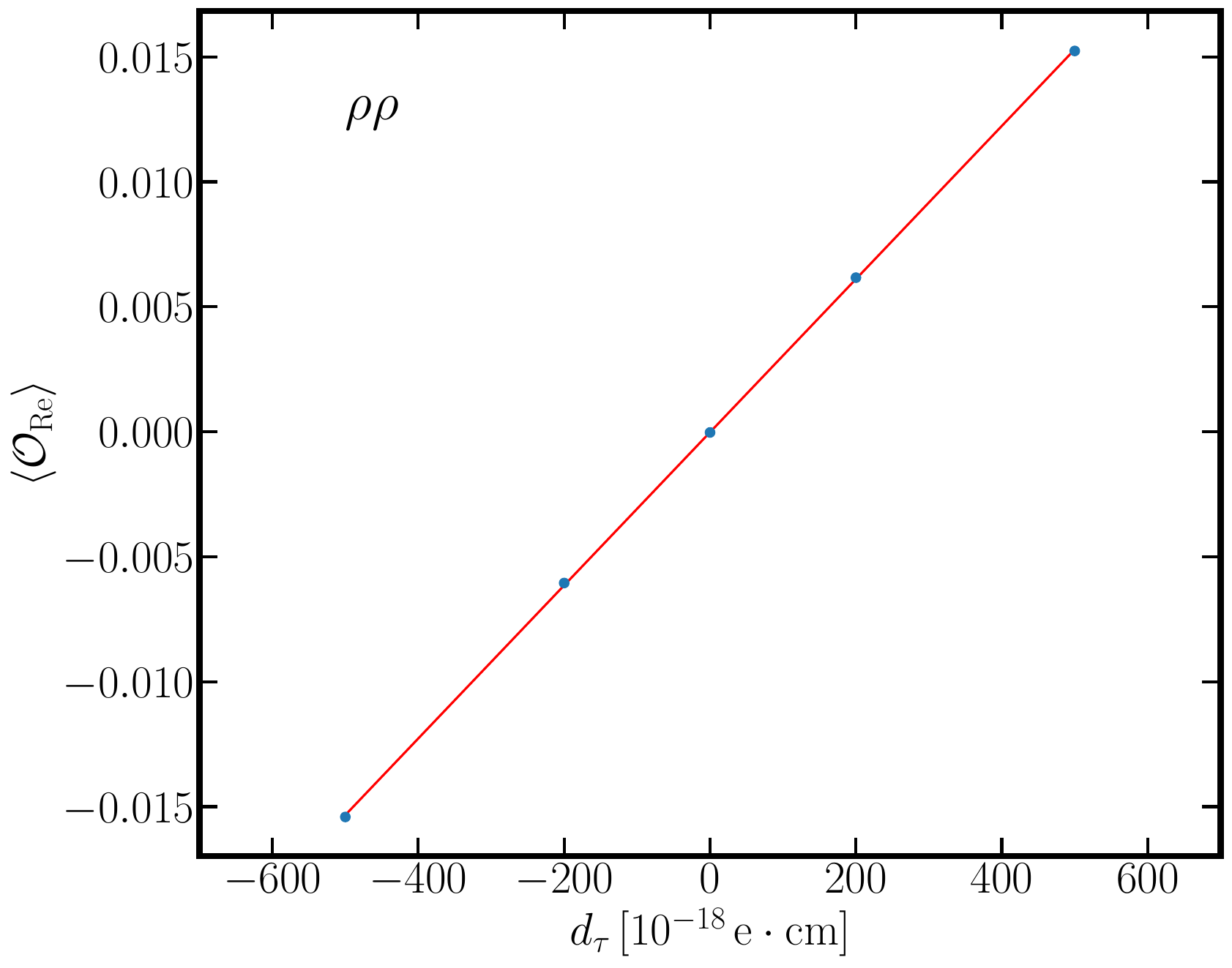}
  \caption{{Left panel: Distribution of $\mathcal{O}_{\operatorname{Re}}$ for different $d_{\tau}$ (Red: $d_{\tau}<0$; Green: $d_{\tau}>0$). The ratio $r$ is defined as the value of $\frac{1}{\sigma}\frac{d\sigma}{d\mathcal{O}_{\operatorname{Re}}}$ for nonzero $d_{\tau}$ divided by that for $d_{\tau}=0$. Right panel: Relationship between $\left<\mathcal{O}_{\operatorname{Re}}\right>$ and $d_{\tau}$.}}
  \label{fig:OO_rhorho}
\end{figure}

MadGraph~\cite{MadGraph} is used along with custom UFO model files to simulate the production of $\tau$ lepton pairs for different values of $d_\tau$. Following the reconstruction of the final states in previous sections, a fast simulation of the detector response is performed with Delphes~\cite{Delphes}. {The distributions of the optimal observables for several different choices of $d_\tau$ from the $\rho\rho$ channel are shown in the left panel of Fig.~\ref{fig:OO_rhorho}. The mean values of the optimal observables are calculated and a linear fit is performed based on Eq.~(\ref{O_d}), which is shown in the right panel of Fig.~\ref{fig:OO_rhorho}. The corresponding result of the linear fit is given in Table~\ref{tab:results_linear_fit} together with the estimated error on the measurement of the mean value of the optimal observables. The estimated sensitivity after 10 years of operation at the STCF is $|d_\tau| < 3.89\times 10^{-18}\,e\cdot\mathrm{cm}$ at a $68\%$ confidence level.} This provides a basis for estimating the EDM of the $\tau$ lepton in future experiments at STCF.

\begin{table}[htbp]
    \centering
    \caption{{The result of the linear fit on the mean values of the optimal observables.}}
    \begin{tabular}{|c|c|c|c|}
    \hline
    Mode   & $a_{\rm Re}\,[(10^{-18}{e\cdot \rm cm})^{-1}]$ & $b_{\rm Re}$  & $\delta\langle \mathcal{O}_{\rm Re}\rangle$\\
    \hline
    $\rho\rho$  & $3.06\times10^{-5}\pm1.43\times 10^{-7}$  & $-1.85\times10^{-5}\pm 4.87\times 10^{-5}$ & $1.088\times 10^{-4}$\\
    \hline
    \end{tabular}
    \label{tab:results_linear_fit}
\end{table}

\section{Summary}
In this study, the process of $\tau$ lepton pair production in $e^+e^-$ collisions was simulated using Monte Carlo methods. Signal photons and signal events were effectively selected through machine learning techniques, significantly improving the signal-to-noise ratio. Based on the simulation data, the event selection algorithm was optimized and its parameters adjusted, resulting in a final signal event selection efficiency of approximately $6.3\%$, with the signal purity increased from $6.2\%$ to $80.0\%$. Regarding particle pairing selection, four methods were compared in terms of their principles and results. A novel approach using joint kinematic fitting was introduced, which greatly enhanced the accuracy of particle pairing, achieving a correct pairing rate of $82.5\%$ for signal events. Additionally, analytical computation was employed to derive the solutions for the $\tau$ lepton momentum, leading to a more precise determination of the number of solutions and their numerical accuracy. Subsequently, the polarimeter vector of the $\tau$ lepton and the squared spin density matrix were calculated. The optimal observables and their relationship with the EDM were obtained, with the estimated sensitivity of $|d_\tau| < 3.89\times 10^{-18}\,e\cdot\mathrm{cm}$ {at $68\%$ confidence level}, laying the groundwork for determining the $\tau$ EDM from experimental data in future STCF experiments.

\section*{Acknowledgements}
We thank Zhipeng Xie, Yupeng Pei, Zekun Jia, Bo Wang, and Mingyi Liu for their valuable discussions. This work is supported by the Supercomputing Center of the University of Science and Technology of China and Lanzhou University. We also thank the Hefei Comprehensive National Science Center for their strong support on the STCF key technology research project.

\appendix
\section{Alternative Particle Pairing Methods}
\label{app:pairing}

\subsection{Method II: Stepwise Kinematic Fitting}
This method has two steps. First, $\gamma_{(1)},\gamma_{(2)}$ are paired into $\pi^0_{(1)}$, and $\gamma_{(3)},\gamma_{(4)}$ into $\pi^0_{(2)}$, followed by a kinematic fitting (using the constraints in Eqs.~(\ref{gamma}) and (\ref{pi0})). Considering the equivalence of $\pi^0_{(1)}$ and $\pi^0_{(2)}$, there are $C_n^2 C_{n-2}^2/2$ possible pairings. The algorithm traverses all pairings and selects the one with the smallest sum of $\chi^2$ from the kinematic fitting. Second, $\pi^0_{(1)}$ is paired with $\pi^+$ and $\pi^0_{(2)}$ with $\pi^-$ for another kinematic fitting (with the constraints in Eqs.~(\ref{pETotal}), (\ref{nu}), (\ref{pipm}), and (\ref{tau})); the reverse pairings are also attempted, with the fit yielding the smaller $\chi^2$ being selected (while still requiring that the neutrino momentum satisfies Eq.~(\ref{pnu_cut})). The particle pairing accuracy in this method is lower than Method I (joint kinematic fitting) due to the two-step kinematic fitting process, where separate $\chi^2$ determinations lead to a sequence of locally optimal choices that may not be globally optimal.

\subsection{Method III: Kinematic Fitting with Resonance Mass of $\rho$}
In $\tau$ decays ($\tau^+ \rightarrow \pi^+ \pi^0 \nu, \tau^- \rightarrow \pi^- \pi^0 \nu, \pi^0 \rightarrow 2\gamma$), $\pi^\pm$ and $\pi^0$ usually form a $\rho$ resonance~\cite{PDG}, allowing the inclusion of the $\rho$ mass constraint:
\begin{equation}
  \begin{aligned}
    \left(E_{\pi^+}+E_{\pi^0_{(1)}}\right)^2 &= \left(\boldsymbol{p}_{\pi^+}+\boldsymbol{p}_{\pi^0_{(1)}}\right)^2 c^2 + m_{\rho}^2 c^4,\\
    \left(E_{\pi^-}+E_{\pi^0_{(2)}}\right)^2 &= \left(\boldsymbol{p}_{\pi^-}+\boldsymbol{p}_{\pi^0_{(2)}}\right)^2 c^2 + m_{\rho}^2 c^4.
  \end{aligned}
  \label{rho}
\end{equation}

Given the broad mass range of the $\rho$ resonance (indicating a large uncertainty), directly adding this constraint to the kinematic fitting would force the fit to adjust physical quantities to strictly satisfy the constraint, which is not desirable. Instead, using the $\pi^0_{(1,2)}$ energies and momenta obtained from the kinematic fitting in the second step of Method II, one can calculate the corresponding $\rho$ masses using Eq.~(\ref{rho}), compare them to the theoretical mass, and select the pairing with the smallest deviation. Using this method, the particle pairing accuracy is still lower than Method I.

\subsection{Method IV: Kinematic Fitting with Momentum Direction}
In this method, after performing the kinematic fitting for all possible pairings, the neutrino momenta $\boldsymbol{p}_{\nu_{(1,2)}}$ are obtained, and the $\tau$ lepton momenta are calculated as $\boldsymbol{p}_{\tau^+} = \boldsymbol{p}_{\pi^+} + \boldsymbol{p}_{\gamma_{(1)}} + \boldsymbol{p}_{\gamma_{(2)}} + \boldsymbol{p}_{\nu_{(1)}}$ and $\boldsymbol{p}_{\tau^-} = \boldsymbol{p}_{\pi^-} + \boldsymbol{p}_{\gamma_{(3)}} + \boldsymbol{p}_{\gamma_{(4)}} + \boldsymbol{p}_{\nu_{(2)}}$. For the correct pairing, the angle between the momenta of the final-state particles and the $\tau$ lepton momenta (considering only the acute angle) should be smaller. Let $\boldsymbol{n}_T$ denote the unit vector along $\boldsymbol{p}_{\tau^+}$, and define:
\begin{equation}
  T=\frac{\sum_{i}\left|\boldsymbol{p}_{i}\cdot\boldsymbol{n}_{T}\right|}{\sum_{i}\left|\boldsymbol{p}_{i}\right|} \quad \left(i=\pi^+,\pi^-,\gamma_{(1)},\gamma_{(2)},\gamma_{(3)},\gamma_{(4)},\nu_{(1)},\nu_{(2)}\right).
\end{equation}

Larger $T$ values are more likely to correspond to the correct physical scenario~\cite{Belle2019}, so this method selects the pairing with the largest $T$ value as the chosen pairing. Results indicate that this method yields a higher pairing accuracy than Methods II and III, but it is still inferior to Method I.

\end{document}